\documentclass[useAMS,usenatbib]{mn2e}

\usepackage{graphicx}
\usepackage{setspace}
\usepackage{natbib}
\usepackage{color}
\usepackage{amsmath,amssymb}
\usepackage{times}
\usepackage{aas_macros}

\title[Discovery of the ATLAS stream]{Discovery of a cold stellar
  stream in the ATLAS DR1 data}

\author[Koposov et al.]{S. E. Koposov $^{1,2}$
  \thanks{E-mail:koposov@ast.cam.ac.uk}, M. Irwin $^{1}$, V. Belokurov
  $^{1}$, E. Gonzalez-Solares$^{1}$, A. Kupcu Yoldas$^{1}$, \newauthor
  J. Lewis$^{1}$, N. Metcalfe$^{3}$, T. Shanks $^{3}$
  \\ $^{1}$Institute of Astronomy, Madingley Rd, Cambridge, CB3 0HA,
  \\ $^{2}$Sternberg Astronomical Institute, Moscow State University,
  Universitetskiy pr. 13, Moscow 119991, Russia,\\ $^{3}$Department of
  Physics, Durham University, South Road, Durham DH1 3LE }

\begin{document}

\date{March 2014}
\pagerange{\pageref{firstpage}--\pageref{lastpage}} \pubyear{2013}

\maketitle

\label{firstpage}
\newcommand{\feh}{\rm [Fe/H]}

\begin{abstract}
We report the discovery of a narrow stellar stream crossing the
constellations of Sculptor and Fornax in the Southern celestial
hemisphere. The portion of the stream detected in the Data Release 1
photometry of the ATLAS survey is at least 12 degrees long, while its
width is $\approx$ 0.25 deg. The Color Magnitude Diagram of this halo
sub-structure is consistent with a metal-poor $\feh \lesssim -1.4$
stellar population located at a heliocentric distance of 20 $\pm$ 2
kpc. There are three globular clusters that could tentatively be
associated with the stream: NGC 7006, NGC 7078 (M15) and Pyxis, but 
NGC 7006 and 7078 seem to have proper motions incompatible with the 
stream orbit.
\end{abstract}


\section{Introduction}

The last decade has witnessed the arrival of unprecedented amounts of
high quality imaging data from the Sloan Digital Sky Survey
(SDSS). Thanks to the depth and the exquisite stability of
the SDSS broad-band photometry across thousands of square degrees on
the sky, previously unseen low-level fluctuations in the Galactic
stellar density field have been unearthed \citep[see
  e.g.][]{NewbergLumps, Willman2005, BelokurovFOS,
  BelokurovReview}. In the Milky Way halo one striking example of a
small-scale over-density is the GD-1 stellar stream \citep{GD1}, only
a fraction of a degree in width, but running over 60 degrees from
end to end in the SDSS footprint. Such stellar trails are formed in
the process of a satellite disruption in the tidal field of the
Galaxy. The mechanics of the stream formation and the subsequent dynamical 
evolution in the host potential have been carefully studied
\citep[e.g.][]{EyreBinney,SandersBinney}. The consensus that has
emerged from both theoretical considerations and admittedly, very
few tests on the actual data \citep[e.g.][]{KoposovGD1}, is that these
structures can be used to yield powerful, unbiased constraints of the
matter distribution in the Galaxy.

In principle there exists a methodology to model the entire spectrum
of tidal debris in the Galactic halo \citep[see e.g.][]{HelmiWhite},
from very narrow structures like GD-1 to broad luminous streams like
that of Sgr \citep[exposed in e.g.][]{MajewskiSgr}, including the
stellar over-densities that do not necessarily even trace out a stream
\citep[e.g.][]{Rewinder}. However, the thinnest streams appear doubly
interesting. First, given that these are not affected by the
progenitor's gravity, methods are now in place to infer the Galactic potential
without the need to resort to approximating the
stellar tracks with a single orbit
\citep[e.g.][]{Bovy2014,Sanders2014}. Second, along these narrow
tidal tails it is easiest to observe density fluctuations due to
interactions with dark matter sub-halos in the Galaxy \citep[see
  e.g.][]{Carlberg2009,Yoon2011}, provided that so-called epicyclic feathering is taken
care of \citep[see e.g.][]{Mastro}. SDSS data has already
been thoroughly mined to yield a handful of cold stellar streams.
These include, for example, the tails of the Pal 5 globular
cluster \citep{Od2001,Od2003,GrillmairPal5}, as well as tails around
NGC 5466 \citep{Belokurov5466}, NGC 5053 \citet{Lauchner2006}, Pal 14
\citep{Sollima2011} and Pal 1 \citep{NO2010}. In addition \citet{GrillmairStyx}
found a group of narrow streams Acheron, Cocytos, Lethe 
and more recently the discovery of the Pisces Stellar Stream was
announced \citep{Bonaca2012, Martin2013}.

\begin{figure*}
  \centering
  \includegraphics[width=0.49\linewidth]{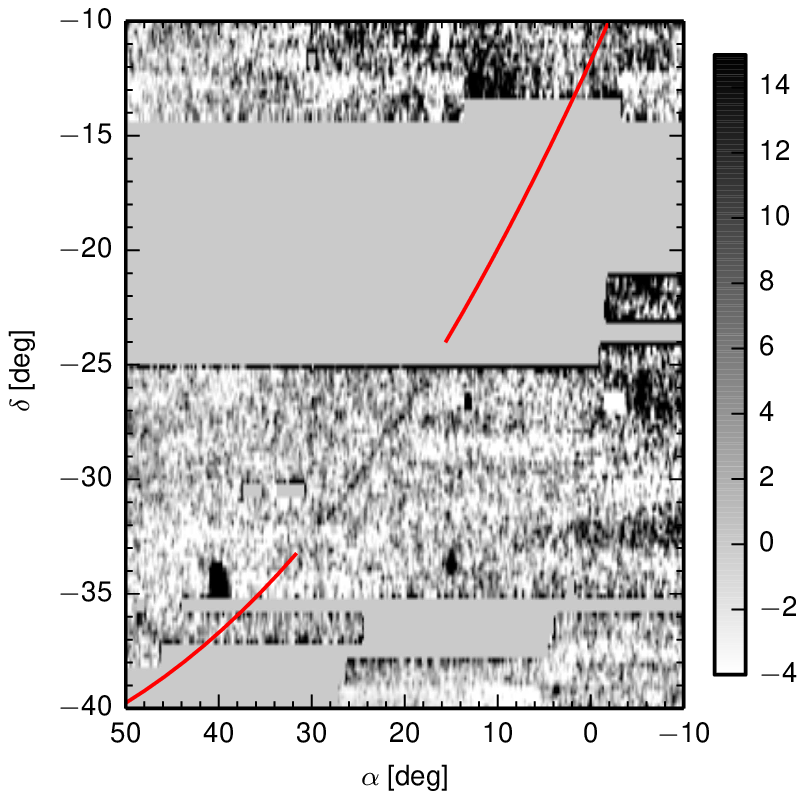}
  \includegraphics[width=0.49\linewidth]{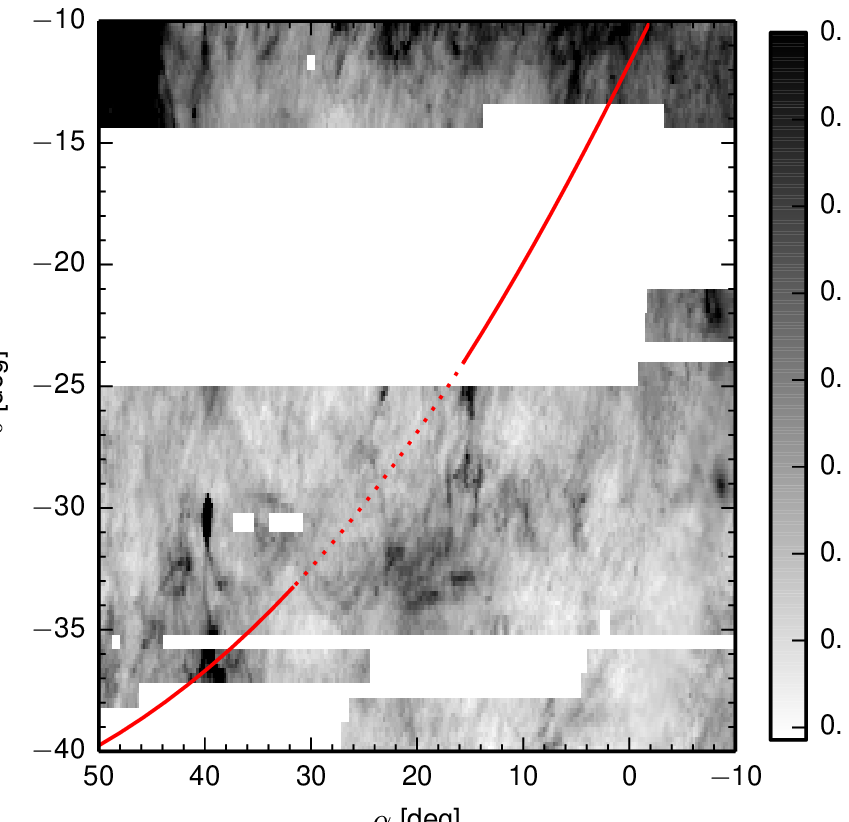}
  \caption[]{\small \textit{Left:} Background subtracted density map of stars optimally weighted 
    by proximity to an isochrone with $\feh =-2.1$ and an age of $12.5$ Gyr
    at a distance of 20 kpc. Darker shades of grey correspond to
    enhanced stellar densities. A narrow stellar stream is clearly
    visible crossing the area diagonally from $\alpha \sim 15^{\circ}$ to
    $\alpha=30^{\circ}$. The red line shows the great circle with
    the pole at $(\alpha,\delta)=(77 \fdg 16, 46 \fdg 92)$ aligned with the
    stream.  The Fornax and Sculptor dwarf spheroidals are visible at
    $(\alpha,\delta)\approx (40^\circ, -35^\circ)$ and $(15^\circ, -33^\circ)$
    respectively while much of the darker shading to the north and west is
    due to the southern Sgr stream.
 \textit{Right:} A map of the Galactic dust extinction
    $E(B-V$) around the stream area \citep{Schlegel} does not seem to reveal 
    any features coincident with the stream.}
   \label{fig:map}
\end{figure*}

In this paper we present the discovery of a new stellar stream in the
Southern celestial hemipshere based on photometry from Data Release 1
(DR1) of the VST ATLAS survey \citep{Shanks2013}.  ATLAS is one of the
three imaging surveys being currently undertaken within the remit of
the ESO VST Public Surveys Program, the other two being KiDS
\citep{Kids} and VPHAS+ (Drew et al 2014).  The aim of the ATLAS
survey is to obtain photometry in the SDSS $ugriz$ filters down to
$r\sim 22$ for approximately 4,500 square degrees of the southern
sky. The primary motiviation for the survey is to identify large
numbers of $z\lesssim 2$ QSOs and Luminous Red Galaxies out to
redshifts of $z\sim 0.6$ for studies of the cosmological matter power
spectrum.  The ability to use this data to also detect low-level
Galactic Halo stellar sub-structure spanning several hundred square
degree survey fields is testament to the quality and the stability of
the ATLAS photometric and astrometric calibration. Section 2 briefly
describes the data being used, while Section 3 gives the details of
the newly discovered stream.  The final section summarises the main
results.

\section{VST ATLAS DR1}
\label{sec:data}

The ATLAS survey makes use of the VLT Survey Telescope, VST, a 2.6\,m ESO 
telescope on Paranal. Images are taken with the OmegaCAM camera mounted at
the Cassegrain focus of the VST.  The camers consists of 32
individual 4k $\times$ 2k CCDs with a pixel size of $\sim$0.21\,arcsec, 
providing a field of view of 1 square degree. Compared to the SDSS, total
exposure times are slightly longer, to account for a smaller pixel size, 
namely 120\,s in $u$, 100\,s in $g$, and 90\,s in each of $riz$. The specified survey
seeing is $<1.4$\,arcsec and the actual median seeing of the ATLAS DR1
is just under 1\,arcsec. The resulting median limiting magnitudes in
each of the five bands corresponding to 5$\sigma$ source detection limits
are approximately $21.0, 23.1, 22.4, 21.4, 20.2$.

The raw data are automatically transferred to the Cambridge Astronomical
Survey Unit (CASU) for further quality checks and subsequent processing. 
CASU processes VST data on a nightly basis including all three of the VST 
public surveys together with any calibration data. The
processing sequence is similar to that used for the IPHAS survey of
the northern Galactic Plane (e.g. Gonzalez-Solares et al 2008),
however the higher level control software is based on that developed
for the VISTA Data Flow System (VDFS, Irwin et al 2004). Science
images are first de-biased and flat-fielded (using series of twilight
sky flats) with particular care taken of 4 detectors suffering from
inter-detector cross-talk. The flatfield sequences plus bad pixel
masks are used to generate confidence maps (e.g. Irwin et al 2004) for
subsequent use in stacking and cataloguing.  The integration time for each
band for ATLAS data is split into two dithered exposures which are stacked
in the pipeline based on derived individual image World Coordinate System 
(WCS) solutions and then catalogued.

Catalogue generation is based on IMCORE\footnote{Software
publicly available from http://casu.ast.cam.ac.uk} (Irwin, 1985) and
makes direct use of the aforementioned confidence maps to downweight
unreliable parts of the images. At this stage, objects are detected,
parameterized and classified morphologically. With catalogues in hand,
the World Coordinate System (WCS) solution based on the telescope
pointing and general system characteristics is progressively refined
using matches between detected objects and the 2MASS catalogue
(Skrutskie et al 2006). The source photometry is calibrated in two
stages. A first-pass solution is obtained using the observations of
standard star fields (observed each night). This is then refined and
adjusted to correct for the scattered light systematics using APASS
all-sky photometric $g,r,i$ catalogues
(http://www.aavso.org/apass). The ATLAS DR1 VST photometry is only in Vega
magnitudes\footnote{The transformation between Vega and AB systems can
  be found in \citet{Blanton2007}}.  Subsequent releases will provide an
independent AB magnitude calibration based on the APASS photometry.
For the analysis that follows we use the Vega system and correct
the magnitudes for Galactic extinction using the dust maps of
\citet{Schlegel}. 

\begin{figure*}
  \centering
  \includegraphics[width=0.49\linewidth]{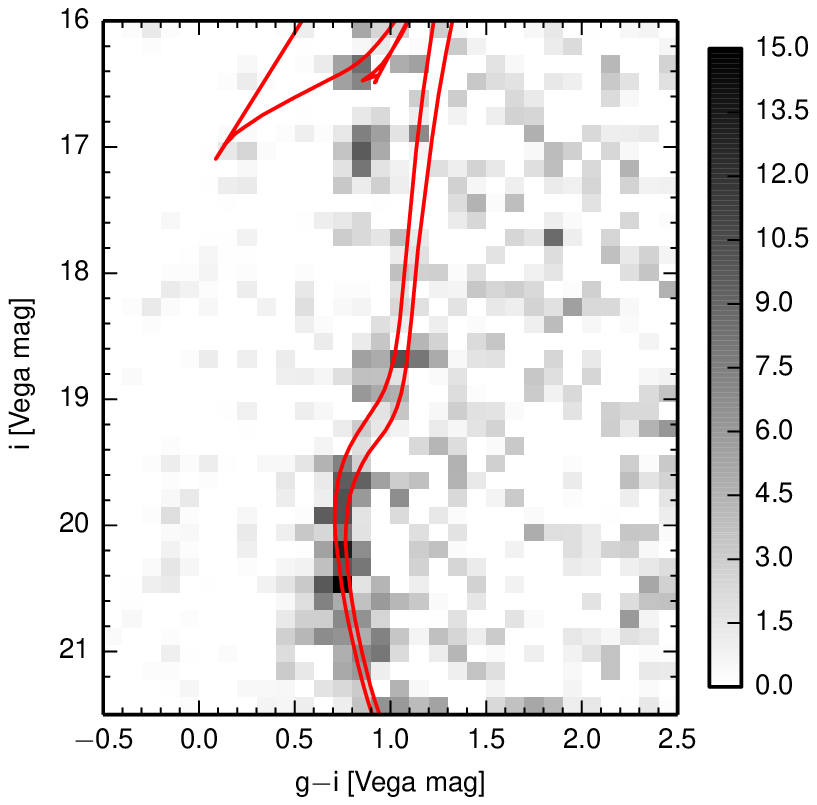}
  \includegraphics[width=0.49\linewidth]{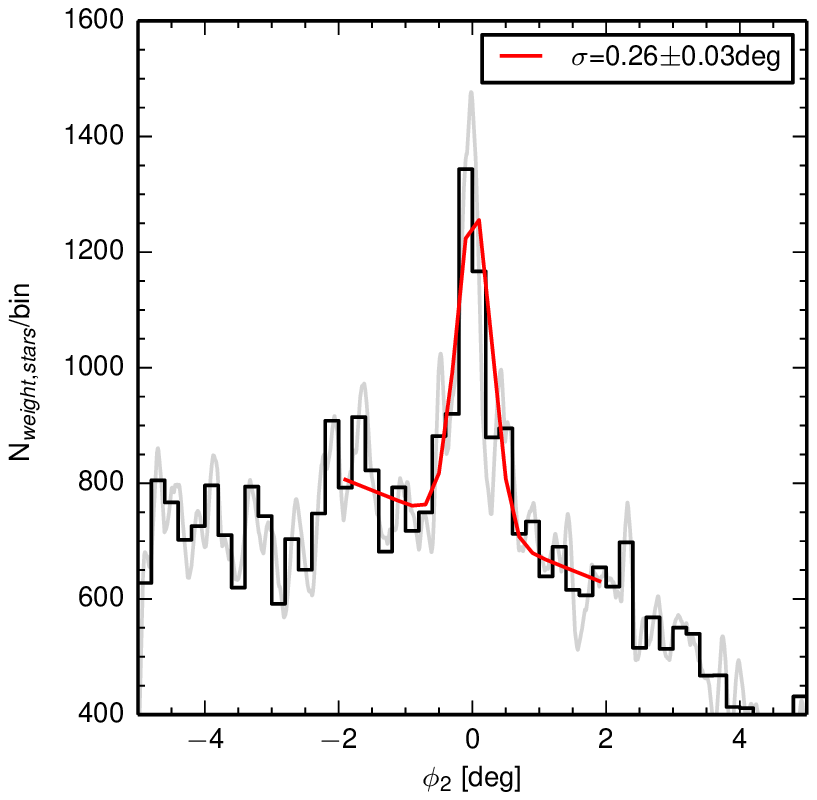}
  \caption[]{\small \textit{Left:} Extinction-corrected g$-$i, i Hess
    diagram of the stream in the range $-3\degr<\phi_1<8\degr$ obtained by
    weighted background subtraction. The magnitudes are in the Vega
    system. The two red lines show isochrones from the \citet{Girardi}
    synthetic library with $\feh = -2.1$ and $-1.45$ (for an age of 12.5
    Gyr) at a distance of 20 kpc. Based on the current photometry
    alone, it is not possible to assign metallicity, age and distance
    to the stream robustly. \textit{Right:} The profile of the stream
    across the range of $-3\degr<\phi_1<8\degr$ obtained using the same optimal
    isochrone weighting as for Fig~\ref{fig:map}. The stream appears as a
    highly significant overdensity on top of the background and can be
    approximately described by a Gaussian with a width (sigma) of $\sim$0.25
    degrees. Grey line shows Kernel Density Estimation with
    Epanechnikov kernel.}
   \label{fig:cmd}
\end{figure*}

\section{The ATLAS Stream}

To test the quality of the ATLAS photometry across a range of colors
and magnitudes, density maps of stars close to the expected location of 
halo main sequence turnoffs at various distances were produced.  As reported 
previously \citep{Shanks2013}, such maps reveal
a smooth slowly varying stellar density field with several
overdensities superposed. Most of this sub-structure is well known, for example,
the southern Sgr stream and the Fornax and Sculptor dwarf spheroidals, however,
a narrow, approximately 12 degree long stream of
stars centered around $(\alpha,\delta)=25^{\circ}, -30^{\circ}$ was immediately
apparent.

Figure \ref{fig:map} shows an enhanced version of the density distribution
of main sequence stars of an area of the sky around the discovered stream. 
This particular map was obtained using the so-called ``matched-filter'' 
technique \citep[e.g.][]{MatchedFilter}. A matched-filter assigns higher
weights to stars in the regions of the Color-Magnitude Diagram (CMD)
dominated by the target stellar population of chosen age, metallicity
and distance, and least contaminated by foreground stars.
These weights, obtained from the ratios of CMD densities
of the target population to the background, are then summed in pixels
on the celestial sphere to expose structures whose stellar populations
are similar to the target. In practice we assume that the stream
stellar population is old and metal-poor and only search for an
optimal distance match. Therefore, a model isochrone with
$\feh=-2.1$ and 12.5\,Gyr age is chosen from the synthetic library
produced by \citet{Girardi}; the CMD tests described below show that
this simplifying assumption is reasonable. The selected isochrone is
placed at different trial distance moduli, the matched-filter density
map is produced and the signal-to-noise ratio of the stream is
estimated. The best signal-to-noise is obtained for a heliocentric distance of
20\,kpc - the distance used to produce the final map shown in the left
panel of Figure~\ref{fig:map}. In order to enhance the contrast of the small scale structures
in the map  we fitted the large scale stellar distribution by 2-dimensional 6-th order polynomial
and subtracted it. The right panel of Figure~\ref{fig:map} shows the
distribution of the Galactic extinction in the same area as the stream.
There is no evidence in the extinction map of any structures related to the dust distribution that
are coincident with the stream.

To analyze the properties of the newly discovered sub-structure we
switch from equatorial coordinates to a system aligned with the
stream great circle. This is achieved by rotating the equatorial coordinate 
pole to a new position at $(\alpha,\delta)=(77 \fdg 16, 46 \fdg 92)$ to give the
along-stream and the across-stream coordinates $\phi_1$ and $\phi_2$
\citep[see e.g.][]{KoposovGD1}; the origin of the rotated coordinates is 
defined to be near the center of the observed part of the stream 
$(\alpha,\delta)=(20 \fdg 0, -26 \fdg 8902)$. The left panel of
Figure~\ref{fig:cmd} shows the Hess difference, that is the CMD density
difference between the region of the sky dominated by the stream
stars, namely $-3^{\circ} < \phi_1 < 8^{\circ}$ and the Galactic
background. Fainter than $i \sim 18$, there are several familiar
features visible, including a sub-giant branch and a main sequence
turn-off. Brighter than $i \sim 17.5$ the Hess difference reveals
several artefacts of imperfect background subtraction. While the
isochrone with metallicity $\feh=-2.1$ seems to describe the CMD of
the stream reasonably well, so does the more metal rich-one, with
$\feh=-1.45$. The $\chi^2$ fit of the Hess diagram by the
$\feh=-2.1$ stellar population reveals the best fit distance of 20\,kpc
and the formal error of 2\,kpc with the expected larger systematic error 
associated with the lack of knowledge of metallicity and age of the stellar
population in the stream.
We conclude that while the assumption of an old and
metal-poor stellar population in the stream is not inadequate, based
on the currently available photometry alone it is not possible to assign
metallicity, age and distance to the stream more robustly. Furthermore 
the existing photometry and the shortness of the visible part of the stream 
does not seem to allow us to constrain the distance gradient along the stream.

\begin{figure*}
  \centering
  \includegraphics[width=0.96\linewidth]{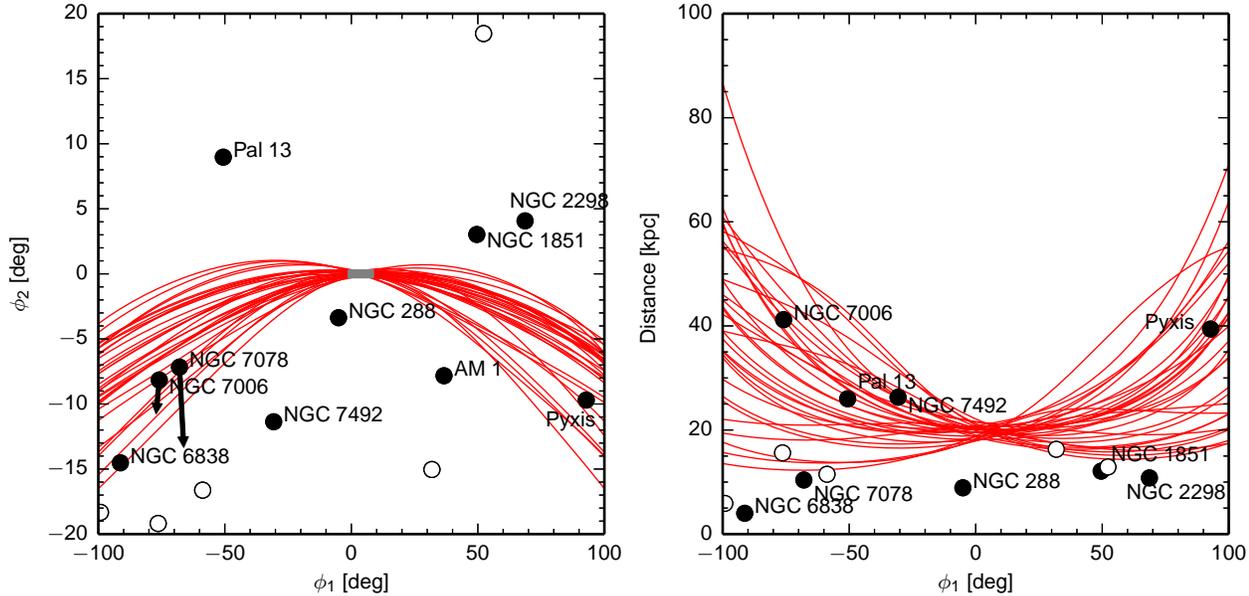}
  \caption[]{\small 30 plausible orbits passing through the detected
    portion of the stream. \textit{Left:} Orbit distribution on the
    sky, in $\phi_1, \phi_2$ coordinates, aligned with the stream. Circles mark the location of
    known globular clusters. Filled (empty) circles represent clusters
    with the across-stream coordinate $\phi_2$ within (outside) the
    range $-15^{\circ} < \phi_2 < 10^{\circ}$. The grey band 
	near $\phi_1\sim$ 0 shows the extend and width of the observed part of
    the stream. \textit{Right:}
    Heliocentric distance change along the test orbits as a function
    of the along-stream coordinate $\phi_1$. The only 3 globular
    clusters within 90$\degr$ of the stream that could possibly lie on orbits passing through the
    stream piece detected are NGC~7006, NGC~7078 and Pyxis. Note
    however that the measured proper motions of NGC~7006 and NGC~7078,
    indicated with black arrows, seem inconsistent with the predicted
    orbital motion.}
   \label{fig:gcs}
\end{figure*}

The right panel of Figure~\ref{fig:cmd} shows the stellar density
distribution across the stream for a range of $-3^{\circ} < \phi_1 <
8^{\circ}$, confirming that the stream detection is highly significant. 
Additionally to the density computed by making the histogram of stars we 
show the density measured by the Kernel Density Estimation using the
Epanechnikov kernel \citep{epanechnikov, wand}, as this density measurement
is invariant to the location of the bin edges of the histogram. 
Also shown is a Gaussian model fit with width $\sigma\approx 0
\fdg 25$. At the distance of 20\,kpc the stream angular width
corresponds to $\sim 90$\,pc. Similarly modest thickness is observed in
several previously detected cold streams that are either known to
originate from star clusters like that of Pal~5 or are believed to be
so, for example GD-1. In order to calculate the total luminosity of the
stream we have calculated the number of stream stars in the
color-magnitude box of $0.1 < (g-i) <0.9$ and $18<i<21$ by fitting a
Gaussian model to the across-stream profile (see
Figure~\ref{fig:cmd}). Within the region of
$-3^{\circ}<\phi_1<8^{\circ}$, the total number of stars $\sim
1200\pm140$. According to the isochrone with the age of 12.5\,Gyr
$\feh=-2.1$ and Chabrier initial mass function, this would
correspond to a total stellar mass $\sim 4 \times 10^4
M_{\odot}$ and luminosity around $M_V \sim -6$. The corresponding 
average surface brightness of the stream is estimated to be $\sim 28-29$ mag/arcsec$^2$.

The question of the likely stream progenitor can be addressed by
exploring the possibility that the stream is produced through the
disruption of one of the known Galactic globular
clusters. Figure~\ref{fig:gcs} displays a representative sample of 30
random possible orbits that pass through the ends of the
stream. i.e. points with $(\phi_1,\phi_2)=(-1,0)$, 
$(\phi_1,\phi_2)=(8.5,0)$,  in a realistic Galactic potential. 
In this experiment, it is assumed
that at $\phi_1=4^{\circ}$ the distance modulus of the stream debris
$m-M = 16.5 \pm 0.1$. The orbits have been obtained by sampling the posterior 
on orbital parameters given the distance and positional constraints, broad prior on radial velocity,
and limit of 100 kpc on the apocenter (to avoid unbound or extremely excentric orbits). 
The gravitational potential of the Galaxy
employed here is identical to that described in
\citet{Fellhauer,KoposovGD1}. The left panel of the figure shows the distribution of
the possible stream orbits on the sky in the $\phi_1, \phi_2$ coordinate system,
while the right panel presents the change of heliocentric distance
along each orbit. From comparing the two panels of the figure it is
clear that within $\sim$ 90 degrees of the stream there are only 3 globular
clusters whose three-dimensional positions could
be coincident with an orbit passing through the portion of the stream
observed in ATLAS DR1. These three are NGC~7006, NGC~7078 (M15) and
Pyxis. The proper motions of the first two globulars have been measured and
therefore can be compared to the model orbit predictions at the
cluster location. According to \citet{m15pm}, the components of the
proper motion of NGC~7078 are $(\mu_{\alpha}, \mu_{\delta})=
(-0.6,-4.0) \pm (0.4, 0.8)$ mas/yr. In the stream coordinate system,
and after removing the contribution of the Solar motion in the Galaxy,
this corresponds to $(\mu_{\phi_1}, \mu_{\phi_2}) = (0.59, -2.21) \pm
(0.6, 0.68)$ mas/yr. Similarly, for NGC~7006, \citet{7006pm} give
$(\mu_{\alpha}, \mu_{\delta}) = (-0.96, -1.14) \pm (0.35, 0.4)$
mas/yr, or $(\mu_{\phi_1}, \mu_{\phi_2}) = (-0.49, -1.00) \pm (0.37,
0.38)$ mas/yr. As is clear from the left panel of Figure~\ref{fig:gcs}
these measured proper motions are inconsistent with the predicted
orbits giving significant non-zero across-stream components
$\mu_{\phi_2}$. But we also note that in the case of NGC~7078 and NGC~7006 the significance 
of the across-stream proper motions is $\sim$ 3.2 $\sigma$ and 2.6 $\sigma$
respectively, so the inconsistency of the stream and proper motion
alignment is only marginally significant.

If the reported proper motions of NGC~7006 and NGC~7078 are correct,
the only globular cluster that can not be ruled out as the stream
progenitor is Pyxis \citep{PyxisIrwin}. The spectroscopic metallicity
of Pyxis is $\feh=-1.4$ \citep{PyxisFeh}, which is consistent with the,
admittedly noisy, CMD presented here. Pyxis stands out somewhat when
compared to the rest of the Galactic globular clusters. First, at 40 kpc 
it falls in between the inner and the outer halo globulars. Second, it is 1$-$2 Gyr
younger that the globulars of similar metallicity \citep{PyxisAge}. 
Finally, Pyxis is very sparse, much like similarly puny Palomar clusters.
Perhaps Pyxis was the first example of an ultra-faint satellite discovered
given its size (in excess of 10 pc) and luminosity ($M_V \sim -5$). We remark also that
the velocity dispersion of Pyxis has been measured using 6 stars \citep{PyxisFeh} and is in the range between 2$-$6\,km/s (1
$\sigma$). While it is entirely possible that the
juxtoposition of one of the tentative stream progenitor orbits and the
three-dimensional position of Pyxis is accidental, wide-angle deep imaging
observations of this peculiar globular are warranted. 

\section{Conclusions}

We have presented the discovery of a new narrow stream in the VST
ATLAS DR1 data. The portion of the stream detected has the following
properties.

\smallskip
\noindent
(1) The stream lies on a great circle with a celestial pole at
$(\alpha,\delta)=77 \fdg 16, 46 \fdg 92$.

\smallskip
\noindent
(2) The heliocentric distance to the stream is $\sim 20$ kpc.

\smallskip
\noindent
(3) The width of the debris distribution on the sky is 0.25 degrees
which at the distance of the stream corresponds to $\sim$ 90 pc
physical size.

\smallskip
\noindent
(4) The CMD of the stream appears to be well described by an old and
metal-poor isochrone. But in order to pinpoint the precise metallicity and age of
 the stream we will need spectroscopic measurements of the
stream members.

\smallskip
\noindent
(5) There are 3 Galactic globular clusters that could plausibly act as
stream progenitors, based on their proximity to the stream orbit: NGC 7006 , NGC 7078
(M15) and Pyxis. 
However, given the proper motions available in the literature, NGC~7006 and NGC~7078 are unlikely to be associated with the stream. It is more difficult to rule out Pyxis at this stage.

\section*{Acknowledgments}
Based on data products from observations made with ESO Telescopes at the La
Silla Paranal Observatory under public survey programme ID programme
177.A-3011(A,B,C). The research leading to these results has received funding from the
European Research Council under the European Union's Seventh Framework
Programme (FP/2007-2013) / ERC Grant Agreement n. 308024. VB
acknowledges financial support from the Royal
Society. S.K. acknowledges financial support from the STFC and the
ERC. We also thank the anonymous referee for thorough review.

\end{document}